\pdfoutput=1
\documentclass[11pt,twoside,a4paper,cmspaper,final,collab]{cms-tdr}
\def\svnVersion{3643d49}\def\svnDate{2025/06/17}\def\cmsCernNoTag{CERN-EP-2024-292}\def\cmsCernDate{\today}\def\cmsMessage{Published in Physical Review D as \href{http://dx.doi.org/10.1103/PhysRevD.111.112004}{\doi{10.1103/PhysRevD.111.112004}.}}
\begin{document}\cmsNoteHeader{EXO-21-016}

\newcommand{\wjets}{\ensuremath{\PW+\text{jets}}\xspace}
\newcommand{\PAGnell}{\HepAntiParticle{\PGn}{\Pell}{}\Xspace} 
\newcommand{\pp}{\ensuremath{{\Pp\Pp}}\xspace}
\newcommand{\taumu}{\ensuremath{\PGt_{\PGm}}\xspace}
\newcommand{\taumutauh}{\ensuremath{\taumu\tauh}\xspace}
\newcommand{\taue}{\ensuremath{\PGt_{\Pe}}\xspace}
\newcommand{\tauetauh}{\ensuremath{\taue\tauh}\xspace}
\newcommand{\taul}{\ensuremath{\PGt_{\Pell}}\xspace}
\newcommand{\taultauh}{\ensuremath{\taul\tauh}\xspace}
\newcommand{\tauhtauh}{\ensuremath{\tauh\tauh}\xspace}
\newcommand{\ptauvis}{\ensuremath{\mathbf{p}^{\PGt\text{vis}}}\xspace}
\newcommand{\pvtauvis}{\ensuremath{\vec{p}^{\,\PGt\text{vis}}}\xspace}
\newcommand{\Etauvis}{\ensuremath{E^{\PGt\text{vis}}}\xspace}
\newcommand{\pZprmiss}{\ensuremath{\mathbf{p}^{\PZpr\text{miss}}}\xspace}
\newcommand{\ptZprmiss}{\ensuremath{\vec{p}_\mathrm{T}^{\,\PZpr\text{miss}}}\xspace}
\newcommand{\zZprmiss}{\ensuremath{p_z^{\PZpr\text{miss}}}\xspace}
\newcommand{\mrec}{\ensuremath{m_{\textrm{rec}}(\PZpr)}\xspace}
\newcommand{\mzp}{\ensuremath{m_{\PZpr}}\xspace}
\newcommand{\leadingl}{\ensuremath{{\Pell}^{\text{l}}}\xspace}
\newcommand{\leadingtauh}{\ensuremath{{\tauh}^{\text{l}}}\xspace}
\newcommand{\BZtautau}{\ensuremath{\mathcal{B}(\PZpr\to\PGtp\PGtm)}\xspace}
\newlength{\cmsFigSpace}
\ifthenelse{\boolean{cms@external}}{\setlength{\cmsFigSpace}{\columnwidth}}{\setlength{\cmsFigSpace}{0.6\textwidth}}

\cmsNoteHeader{EXO-21-016}
\title{Search for heavy neutral resonances decaying to tau lepton pairs in proton-proton collisions at \texorpdfstring{$\sqrt{s} = 13\TeV$}{sqrt(s) = 13 TeV} }

\date{\today}

\abstract{
	A search for heavy neutral gauge bosons (\PZpr) decaying into a pair of tau leptons is performed in proton-proton collisions 
	at $\sqrt{s}=13\TeV$ at the CERN LHC. The data were collected with the CMS detector and correspond to an integrated 
	luminosity of 138\fbinv. The observations are found to be in agreement with the expectation from standard model 
	processes. Limits at 95\% confidence level are set on the product of the \PZpr production cross section and its branching fraction to tau lepton pairs for a range of \PZpr boson masses. 
	For a narrow resonance in the sequential standard model scenario, a \PZpr boson with a mass below 3.5\TeV is excluded. This is the most stringent limit to date from this type of search.
}

\hypersetup{
pdfauthor={CMS Collaboration},
pdftitle={Search for heavy neutral resonances decaying to tau lepton pairs in proton-proton collisions at sqrt(s) = 13 TeV},
pdfsubject={CMS},
pdfkeywords={CMS, heavy resonance, taus, BSM}}

\maketitle

\section{Introduction}
\label{sec:introduction}

The CMS experiment~\cite{CMS:2021ctt} at the CERN LHC is conducting an extensive program of searches for
evidence of new particles predicted by various extensions of the standard model (SM).
One of the simplest extensions postulates an additional $U(1)$ symmetry group that gives rise to the production of a new neutral gauge particle, analogous to the 
\PZ boson in the SM, commonly referred to as a \PZpr boson. There are other 
more complicated extensions of the SM that also predict the existence of similar mediators, such as grand unified theories~\cite{Mohapatra:1985xm}, 
$E(6)$ models~\cite{London:1986dk}, and the Randall--Sundrum model~\cite{Randall:1999ee}. 
Searches for \PZpr bosons have been carried out by the ATLAS and CMS Collaborations, placing stringent bounds on \PZpr boson production as a function of its mass, \mzp, 
especially for the $\PZpr\to\PGmp\PGmm$ and $\Pep\Pem$ decay channels.
Combining the $\PGmp\PGmm$ and $\Pep\Pem$ final states and using the sequential SM (SSM) extension, the  
production of an SSM \PZpr boson has been excluded at 95\% confidence level (\CL) for masses below 5.15\,(5.10)\TeV in searches reported by
CMS~\cite{CMS:2021ctt} (ATLAS~\cite{ATLAS:2019erb}). 
The SSM assumes that the \PZpr boson has the same couplings to SM particles as those of the \PZ boson, in particular, couplings to SM fermions that are independent of their generation. 
There are several models that instead propose preferential couplings of the \PZpr boson to third-generation fermions~\cite{Langacker:2008yv,Lynch:2000md}. 
In models with nonuniversal lepton couplings, assuming a \PZpr boson with the same couplings to third generation fermions as the SM \PZ boson, 
bounds on \mzp are found for the $\PZpr \to \PGtp\PGtm$ decay mode~\cite{ATLAS:2017eiz,CMS:2016xbv} excluding \mzp up to 2.42\TeV and
the $\PZpr \to \PQb\PAQb$ decay mode~\cite{ATLAS:2019fgd,CMS:2022eud} excluding \mzp up to 2.7\TeV.

In recent years, interest in potential effects beyond the SM has emerged as a result of anomalies observed in the precision measurements of \PB meson decay rates,
including $R(\PD^{(*)})$~\cite{BaBar:2012obs,BaBar:2013mob,  Huschle:2015rga,Aaij:2015yra, Sato:2016svk, Hirose:2016wfn, Hirose:2017dxl, Aaij:2017uff,LHCb:2017rln,LHCb:2023zxo,LHCb:2024jll},
the ratio of branching fractions with leptons of different flavors.
The $R(\PD^{(*)})$ ratio is sensitive to the violation of lepton flavor universality in $\PQb \to \PQc \mathrm{l} \PAGnl$ transitions, where $\mathrm{l}$ denotes any lepton flavor.
These anomalies have led to the proposal of a large number of theoretical models that predict the existence of heavy neutral and charged gauge bosons (\PZpr and \PWpr, respectively) 
with nonuniversal fermion couplings within a new $SU(2)$ group. These models further motivate the search for a \PZpr boson that preferentially couples to third-generation fermions~\cite{thirdgeneration,banomalies}. 
For this search, we use a simplified phenomenological model, in which the \PZpr boson mass and couplings are treated as free parameters~\cite{Florez:2016uob}. 
In this paper, we consider a \PZpr boson with the same fermion couplings as the \PZ boson to compare with previous searches, but also consider scenarios with varied \PZpr boson couplings to the \PGt leptons.

This paper presents a search for a \PZpr boson via the $\PGtp\PGtm$ decays, considering production via quark-antiquark annihilation, in proton-proton (\pp) collisions 
at $\sqrt{s} = 13\TeV$ at the LHC. 
The data were collected with the CMS experiment during 2016--2018, and correspond to an integrated luminosity of 138\fbinv.  Tabulated data are provided in the HEPData record~\cite{hepdata} for this analysis.

The remaining sections of this paper present
the analysis strategy, in Section~\ref{sec:strategy}, followed by a short detector description in Section~\ref{sec:detector}. 
The event reconstruction is detailed in Section~\ref{sec:reconstruction} and the generation of simulated samples in Section~\ref{sec:simulation}. 
We describe the event selection criteria and background estimation in Sections~\ref{sec:selection} and~\ref{sec:background}, respectively, for each of the \PGt pair decay channels. 
Taking into account systematic uncertainties as discussed in Section~\ref{sec:systematics}, we present the results in Section~\ref{sec:results} and a summary in Section~\ref{sec:summary}.

\section{Analysis strategy}
\label{sec:strategy}

The \PGt decay channels can be broadly classified into hadronic (\tauh), $\PGnGt+\text{hadrons}$, and leptonic (\taul), $\PGnGt+\Pell+\PAGnell$, with \Pell representing \PGm or \Pe.
The analysis presented here includes the three $\PZpr \to \PGtp \PGtm$ decay channels with the 
largest \PGt decay branching fraction products:  
$\PZpr \to\PGtpm_{\PGm} \tauh^{\mp}$, $\PZpr \to\PGtpm_{\Pe} \tauh^{\mp}$, and $\PZpr \to \tauh^{+} \tauh^{-}$, referred to below as \taumutauh,  \tauetauh, and \tauhtauh, respectively.

In the production of \PZpr bosons via $s$-channel annihilation of quark-antiquark pairs, the \PZpr bosons have limited Lorentz boost in the plane transverse to the beam. 
For the \TeVns-scale \mzp of interest, the \PGt decay daughters tend to be clustered within small back-to-back cones. 
With this picture in mind, we derive an estimate of the \PZpr boson candidate mass \mrec,  making use of the transverse momentum imbalance between the visible daughters of the two \PGt lepton decays:
\ifthenelse{\boolean{cms@external}}
{
\begin{widetext}
\begin{equation}
\mrec = \sqrt{(\Etauvis_{1}+\Etauvis_{2}+\abs{\pZprmiss})^2 - \abs{\ptauvis_{1}+\ptauvis_{2}+\pZprmiss}^2}.
\label{eq:mrecolephad}
\end{equation}
\end{widetext}
}
{
\begin{equation}
\mrec = \sqrt{(\Etauvis_{1}+\Etauvis_{2}+\abs{\pZprmiss})^2 - \abs{\ptauvis_{1}+\ptauvis_{2}+\pZprmiss}^2}.
\label{eq:mrecolephad}
\end{equation}
}
Here $(\Etauvis_i, \ptauvis_i)$ is the four momentum of the reconstructed daughter(s) of the $i^{\mathrm{th}}$ \PGt candidate.  The momentum of the invisible neutrinos is represented by
$\pZprmiss = (\ptZprmiss,\zZprmiss)$, with $\ptZprmiss = -(\pvtauvis_{1\mathrm{T}} + \pvtauvis_{2\mathrm{T}})$ and $\zZprmiss=0$.
We find that this definition of \ptZprmiss provides better discrimination between the signal and the SM background processes than the event momentum imbalance \ptvecmiss, the negative vectorial sum of all reconstructed objects. This is because \ptvecmiss includes contributions from mismeasured jets in the underlying event and from additional \pp interactions within the same or nearby bunch crossings (pileup).

We use \mrec as the primary discriminating variable for extraction of the signal yield.  Although the natural decay width of the \PZpr boson is small compared with its mass 
($<$3\% for the model assumptions in this paper), the relative resolution on \mrec is between 30 and 45\% for the range of \mzp considered. 
The analysis is thus dominated by experimental resolution effects arising from the undetected neutrinos.
Data and simulation are used to obtain estimates of the yields of the SM background processes, and the signal yield is obtained from a maximum likelihood (ML) fit to the \mrec distributions for each of the three $\PZpr\to\PGtp\PGtm$ decay channels, as described in the following sections.

\section{The CMS detector}
\label{sec:detector}

The central feature of the CMS apparatus is a superconducting solenoid of 6\unit{m} internal diameter, providing a magnetic field of 3.8\unit{T}. Within the solenoid volume are a silicon pixel and strip tracker, a lead tungstate crystal electromagnetic calorimeter (ECAL), and a brass and scintillator hadron calorimeter, each composed of a barrel and two endcap sections. Forward calorimeters extend the pseudorapidity ($\eta$) coverage provided by the barrel and endcap detectors. Muons are reconstructed in gas-ionization detectors embedded in the steel flux-return yoke outside the solenoid. More detailed descriptions of the CMS detector, together with a definition of the coordinate system used and the relevant kinematic variables, can be found in Refs.~\cite{CMS:2008xjf,CMS:2023gfb}.

Events of interest are selected using a two-tiered trigger system. The first level, composed of custom hardware processors, uses information from the calorimeters and muon detectors to select events at a rate of 
around 100\unit{kHz} within a fixed latency of about 4\mus~\cite{CMS:2020cmk}. The second level, known as the high-level trigger, consists of a farm of 
processors running a version of the full event reconstruction software optimized for fast processing, and reduces the event rate to around 1\unit{kHz} before data storage~\cite{CMS:2016ngn}. 

\section{Object reconstruction}
\label{sec:reconstruction}

The first stage of particle reconstruction employs the particle-flow (PF) algorithm~\cite{CMS:2017yfk}, which aims to reconstruct and identify each individual visible particle in an event (photon, electron, muon, charged hadron, neutral hadron) with an optimized combination of all subdetector information.  Further identification (ID) strategies for particular particle types are applied to achieve an appropriate balance between efficiency and the purity of the resulting candidate collections.

In the presence of pileup, the primary vertex, at which the \pp collision of interest occurs, is taken to be the vertex corresponding to the hardest scattering in the event, evaluated using tracking information alone, as described in Section~9.4.1 of Ref.~\cite{CMS-TDR-15-02}.

Jets are reconstructed offline from PF candidates using the anti-\kt clustering algorithm~\cite{Cacciari:2008gp, Cacciari:2011ma} with a distance parameter of 0.4.  
Jet momentum is determined as the vectorial sum of all particle momenta in the jet, and is found from simulation to be, on average, within 5 to 10\% of the true momentum over the entire \pt spectrum 
and detector acceptance. Pileup interactions can contribute additional tracks and calorimetric energy depositions, increasing the apparent jet momentum. To mitigate this effect, tracks identified to be originating from pileup vertices are discarded, and an offset correction is applied to correct for remaining contributions~\cite{CMS:2020ebo}. Jet energy corrections are derived from simulation studies so that the average measured energy of jets matches that of particle level jets. In situ measurements of the momentum balance in dijet, $\text{photon} + \text{jet}$, $\PZ + \text{jet}$, and multijet events are used to determine any residual differences between the jet energy scale in data and in simulation, and appropriate corrections are made~\cite{CMS:2016lmd}. Additional selection criteria are applied to each jet to remove jets potentially dominated by instrumental effects or reconstruction failures~\cite{CMS:2020ebo}.

The \ptvecmiss vector is computed as the negative vector sum of the transverse momenta of all the PF candidates in an event, modified to account for corrections to the energy scale of the reconstructed jets in the event~\cite{CMS:2019ctu}.  Its magnitude is denoted \ptmiss.

Muons are measured in the pseudorapidity range $\abs{\eta} < 2.4$, with detection planes made using three technologies: drift tubes, cathode strip chambers, and resistive-plate chambers. 
The single-muon trigger efficiency exceeds 90\% over the full $\eta$ range, and the efficiency to reconstruct and identify muons is greater than 96\%. 
Matching muons to tracks measured in the silicon tracker results in a relative transverse momentum resolution of 1\% in the barrel and 3\% in the endcaps for muons with \pt up to 100\GeV.
The \pt resolution in the barrel is better than 7\% for muons with \pt up to 1\TeV~\cite{CMS:2018rym}.
This analysis uses the tight working point~\cite{CMS:2018rym} of the muon ID. 
To suppress muons from hadron decays and other nonprompt sources, an additional requirement is imposed on the relative isolation, defined as the ratio of the energy from neutral and charged PF candidates in a cone of $\Delta R \equiv \sqrt{\smash[b]{(\Delta\eta)^2+(\Delta\phi)^2}} < 0.4$ around the muon trajectory to that of the muon. 
A correction for energy mismeasurement from pileup is applied, and we require this corrected relative isolation to be smaller than 0.15.

Electrons are reconstructed using energy deposits in the ECAL detector that have a matching track in $(\eta,\,\phi)$ space in the silicon tracking system, 
within $\abs{\eta} < 2.4$. Electron candidates that fall in the transition region between the barrel and the endcap of the CMS detector, 
located at  $1.44 < \abs{\eta} < 1.57$, are not considered. The momentum resolution for electrons with $\pt \approx 45\GeV$ from $\PZ \to \Pe \Pe$ 
decays ranges from 1.6--5.0\%. It is better in the barrel region than in the endcaps, and depends on the bremsstrahlung energy emitted by the electron 
as it traverses the material in front of the ECAL~\cite{CMS:2020xlg,CMS:2020uim}. 
To identify electrons, we use the high-energy electron pairs (HEEP) ID for the selection of signal events;
for veto and control region (CR) selections, we use instead the loose working point of the cut-based electron ID~\cite{CMS:2020uim}. 
Isolation is defined as the amount of energy in a cone of $\Delta R < 0.3$ around the electron direction. 
It is an integral part of both electron ID criteria used, as defined in detail in~\cite{CMS:2020uim}, and both IDs utilize electron-energy-dependent isolation criteria with pileup energy corrections that differ between the barrel and endcap regions of the detector. 

{\tolerance=800 Hadronic \PGt lepton decays are reconstructed from jets using the hadrons-plus-strips algorithm~\cite{CMS:2018jrd}, which combines 1 or 3 tracks with energy deposits in the calorimeters to form \tauh candidates. Neutral pions are reconstructed as strips with dynamic size in $(\eta,\,\phi)$ from reconstructed electrons and photons, where the strip size varies as a function of the \pt of the electron or photon candidate. \par}

The \textsc{DeepTau} deep neural network algorithm~\cite{CMS:2022prd} is used to distinguish genuine \tauh candidates from jets originating from the hadronization of quarks or gluons, and from electrons or muons. 
Information from all individually reconstructed particles near the \tauh momentum direction is combined with properties of the \tauh candidate and of the event.  We make use of the tight (medium) working point for the discrimination of \tauh from muons (electrons), and of the tight, loose, and very loose working points for discrimination against jets, as described in Sections~\ref{sec:selection} and~\ref{sec:background}.

To identify jet candidates from \PQb quark hadronization (hereafter denoted \PQb jets) we use the \textsc{DeepCSV} \PQb tagging algorithm~\cite{BTV-16-002}, which is based on a deep neural network.
We employ the tight working point of the discriminator, for which the typical efficiency is 50\%, while the mistag rate of jets originating from light-flavored quarks or gluons is 0.1\% and from \PQc jets is 2.4\%.

\section{Simulated samples}
\label{sec:simulation}

Monte Carlo simulation is used in the analysis to design the event selection and to model the signal and SM background processes.  For estimates of the background yields we use a combination of simulation and data, as described in Section~\ref{sec:background}.

The background from SM events composed uniquely of jets produced via the strong interaction is referred to as quantum chromodynamics (QCD) multijet events. 
Events from Drell--Yan (DY), QCD multijet, single \PQt quark, and \wjets processes are simulated with the \MGvATNLO 2.6.0 event generator~\cite{Alwall:2014hca} at leading order (LO) accuracy.  
The \POWHEG 2.0~\cite{Nason:2004rx, Frixione:2007vw, Alioli_2010} event generator at next-to-LO (NLO) accuracy in QCD is used to simulate \ttbar events, while diboson ($\PV\PV$) processes are produced at LO with \PYTHIA~8.226 (8.230) for data taken in 2016 (2017 and 2018)~\cite{Sjostrand:2014zea}. 

Signal events are generated using \MGvATNLO at LO~\cite{Fuks:2017vtl}. The influence of interference effects between the \PZpr signal diagrams and the SM processes are found to be small, resulting in no significant impact on the interpretation of the results, for the range of  \PZpr boson masses and couplings considered in this analysis. 
A set of samples is produced for \PZpr boson masses ranging from 250 to 3000\GeV in steps of 250\GeV, as well as at 4000 and 5000\GeV.
The primary set of benchmark signal samples for interpretation of the results have the \PZpr boson couplings set to SM \PZ values. 
Additionally, scenarios with varied modifiers $g_{\PGt}$ of the couplings to \PGt leptons are considered. 
The theoretical decay width of the \PZpr boson in the production model increases with $g_{\PGt}$ but remains smaller than the experimental resolution ($<$30\%) for the range of $g_{\PGt}$ values considered in this study. 

{\tolerance=800 The \MGvATNLO and \POWHEG generators used to simulate signal and background processes are interfaced with \PYTHIA~8.226 (8.230) using the CUETP8M1 (CP5) tunes~\cite{Khachatryan:2015pea,CMS:2019csb} for parton showering and fragmentation in the 2016 (2017--2018) simulated samples. 
The NNPDF3.0 (3.1) NNLO~\cite{Ball:2014uwa} parton distribution functions (PDFs) are used in the event generation for 2016 (2017--2018) samples. 
The cross sections used to normalize the simulated background yields are calculated at next-to-NLO or NLO accuracy~\cite{Alioli:2009je,Re:2010bp,Gavin:2010az,Melia:2011tj,Beneke:2011mq,
Cacciari:2011hy,Baernreuther:2012ws,Czakon:2012zr,Czakon:2012pz,Gavin:2012sy,Czakon:2013goa,Alwall:2014hca}, while those used for the signal yields are calculated at NLO accuracy. \par}

The detector response is modeled with \GEANTfour~\cite{Agostinelli:2002hh}. 
Pileup is incorporated by simulating additional minimum bias interactions with the multiplicity distribution matching that measured in data.

\section{Event selection criteria}
\label{sec:selection}

The signal region (SR) selection criteria for each $\PZpr \to \PGtp \PGtm$ decay channel are chosen to maximize the expected signal significance by balancing signal efficiency 
against the suppression of background.
The main SM background processes are DY, \wjets, \ttbar, QCD multijet, and $\PV\PV$ ($\PW\PW$, $\PZ\PZ$, and $\PW\PZ$) production. 
The DY process constitutes an irreducible background since $\PGtp \PGtm$ pairs arise from the intermediate \PZ and photon states.
However, since the analysis focuses on massive \PZpr bosons, the visible daughters from the two \PGt lepton decays 
typically have higher momentum and larger angular separation than those from DY events. We exploit these characteristics 
to separate heavy \PZpr boson signals from background. Events from \wjets and \ttbar can mimic the \taultauh final states when the \PW boson, produced directly or from a top quark decay, decays via 
$\PW \to \Pell \PGn_{\Pell}$, and a jet is misidentified as a \tauh candidate.  Events from $\PV\PV$ or \ttbar production with leptonic decays of both vector bosons produce genuine lepton pairs, leading to final states with a $\PGtp \PGtm$ pair as in signal events. Multijet events become a background source 
through the misidentification of jets as \tauh or \Pell candidates.

Events are required to have one \tauh candidate accompanied by a \PGm, \Pe, or second \tauh candidate of opposite-sign (OS) charge, all within $\abs{\eta}<2.1$.
To avoid possible overlaps among the three channels, we reject events containing additional lepton candidates of any flavor beyond the selected OS lepton candidate pair.  
Specifically, the additional lepton can be a  \PGm or \Pe candidate with  $\pt > 10\GeV$ and $\abs{\eta} < 2.1$, or a \PGt candidate meeting the \tauh requirements of the particular channel.
Since the analysis explores the production of \PZpr bosons through quark-antiquark annihilation, events from vector boson fusion are suppressed by vetoing events with a pair of jets $j^{1,2}$, each of which passes $\pt > 30\GeV$ and $\abs{\eta} < 4.7$, with a pseudorapidity separation $\abs{\Delta \eta(j^{1}, j^{2})} > 4.2$ and an invariant mass above 500\GeV. 

For the ID of \tauh candidates in the SRs for all decay channels we impose the tight working points of the discriminators against jets and muons, and the medium one against electrons, as defined in Ref.~\cite{CMS:2022prd}.
The probability of misidentifying a jet as a \tauh candidate depends on the \pt and flavor of the jet; for the kinematic conditions of this analysis it has been estimated in a \wjets sample to be approximately 
0.6\%, with a genuine \tauh ID efficiency of 60\%. The probability for muons (electrons) to be misidentified as \tauh candidates is 0.07\% (0.2\%) for a genuine \tauh ID efficiency of $>$99\% (90\%).

\subsection{The \texorpdfstring{\taumutauh}{muonic-hadronic} SR}
\label{sec:selmutauh}

The \taumutauh events are required to satisfy the single-muon trigger, whose efficiency exceeds 90\% over the full $\eta$ range for the selected muon candidates with $\pt > 35\GeV$.  Muons are also required to
be well isolated and to pass the tight ID criteria defined in Ref.~\cite{CMS:2018rym}.  The \tauh candidate is required to satisfy $\pt > 20\GeV$.  The two candidates are required to
be well separated in $(\eta,\,\phi)$ space by the criterion $\Delta R \equiv \sqrt{\smash[b]{(\Delta \phi (\taumu, \tauh))^{2} + (\Delta \eta (\taumu, \tauh))^{2}}} > 0.3$. 
We reject events containing \PQb jet candidates with $\pt > 30\GeV$ and $\abs{\eta} < 2.4$.
Events from DY, \wjets, and QCD multijet production are significantly suppressed by requiring the \taumu and \tauh candidates to have a large azimuthal separation given by $\cos{\Delta\phi(\taumu, \tauh)} < -0.98$.
In addition, we require that \ptvecmiss lie in the direction opposite that of the \tauh or of the \taumu candidate with the highest \pt (leading lepton, \leadingl), by requiring $\cos{\Delta \phi (\ptmiss, \leadingl)} < -0.95$.
This requirement further reduces the contribution of \wjets and QCD multijet events. The transverse mass of the leading lepton 
and \ptvecmiss, $\mT (\ptmiss, \leadingl)= \sqrt{\smash[b]{\ptmiss \pt(\leadingl)(1 - \cos{(\Delta \phi(\ptmiss,\leadingl))})}}$,
is required to be greater than 150\GeV for further suppression of \wjets events.  

\subsection{The \texorpdfstring{\tauetauh}{electronic-hadronic} SR}
\label{sec:seletauh}

Similar selection criteria are applied to the \tauetauh channel, with the following differences.  We require these events to satisfy a single-electron trigger that has an efficiency above 90\% for 
electrons after the requirement $\pt > 35\,(55)\GeV$ for data collected in 2016 (2017--2018).                                                         

\subsection{The \texorpdfstring{\tauhtauh}{dihadronic} SR}
\label{sec:seltauhtauh}

For the \tauhtauh channel we select events that satisfy a dedicated trigger~\cite{CMS:2020cmk} with at least two \tauh candidates. We require each \tauh candidate to have $\pt > 70\GeV$, ensuring a trigger efficiency of at least 90\%.  The two \tauh candidates must be separated by $\Delta R > 0.3$. 
Events with any \PQb jet candidate having $\pt > 30\GeV$ and $\abs{\eta} < 2.4$ are removed, to suppress top quark backgrounds. 
To reduce the contribution of DY events, the reconstructed mass of the \tauh pair is required to exceed 100\GeV. To discriminate against \wjets and QCD multijet events, we require
the two \tauh candidates to have a large azimuthal separation, $\cos{\Delta \phi (\tauh^{1}, \tauh^{2})} < -0.95$, 
while the \ptvecmiss  and the leading-\pt \tauh candidate \leadingtauh are required to satisfy $\abs{\cos{\Delta \phi (\ptmiss, \leadingtauh)}} > 0.9$. Further suppression of the contribution of QCD multijet events is achieved with
a requirement $\ptmiss > 30\GeV$.

\section{Background estimation}
\label{sec:background}

The background yields are estimated in discrete intervals (bins) in the variable \mrec, by
methods that depend on the SM process and the channel. In most cases, these estimates are based on samples of simulated events.
Differences between data and simulation with respect to the efficiency of triggers, particle ID or isolation, and energy scale and resolution of electrons, muons, jets, and \ptmiss are taken into account with dedicated correction factors whose uncertainties are propagated throughout.
These correction factors are described in the references cited in Section~\ref{sec:reconstruction}.
Residual mismodeling can remain due to systematic effects of the theoretical predictions,
or to the phase space requirements specific to this analysis or to a specific channel.
To ensure that the effect of any such mismodeling and its uncertainty in the background yield is accounted for, we evaluate a data-to-simulation scale factor (SF) in a CR.
The systematic uncertainties in the SFs account for correlations between CRs and between the CRs and SRs.
Validation tests are performed using additional simulation and data control samples to ensure the correct determination of the SFs and their applicability to the SR background prediction. 

The SFs are found to be approximately independent of \mrec, and are thus applied to the \mrec distributions as bin-independent normalization factors.
Uncertainties from the residual shape dependence are evaluated as described in Section~\ref{sec:systematics}.  The SFs are derived for each data-taking year separately, along with their uncertainties and year-to-year correlations.  The values of the SFs and their uncertainties are given in the following subsections.  With a few exceptions, these values are consistent with unity.
In the case of the QCD multijet background, the estimate is derived from the yield measured in dedicated orthogonal data control samples using the ABCD method described in Section~\ref{sec:bkgtaultauh}. 
For SM processes that contribute less than 5\% to the background yield, the simulation is taken directly.

Each CR is constructed to be enriched in a given SM process and orthogonal to the corresponding SR, but with characteristics similar to those of the SR.  We correct for contamination from other SM processes by subtracting their SF-corrected simulated yields.  The resulting dependence of the estimate of one background process on the SF for another is treated through the uncertainty correlation terms in the ML fit.

\subsection{The \texorpdfstring{\taultauh}{leptonic-hadronic} CRs}
\label{sec:bkgtaultauh}

The main sources of background events in the \taultauh channels are DY, \wjets, and \ttbar production. 

To derive the SFs for the DY background in the \taultauh channels we define CRs by selecting events as for the SRs except for an inverted \mT criterion, 
$\mT < 150\GeV$, and a requirement that the reconstructed mass of the \taultauh pair lie within the \PZ boson mass window, $70 < m_{\text{rec}}(\taul\tauh) < 110\GeV$. 
Since the higher \pt threshold for electrons of 55\GeV in 2017--2018 would deplete the low end of this mass range for a \tauetauh CR, for those data sets we substitute the SFs measured in the corresponding \taumutauh channels. 
The values of the DY SFs measured in the \taumutauh channel are $0.98\pm0.06$, $0.91\pm0.05$, and $0.90\pm0.06$ in 2016, 2017, and 2018, respectively.
The value of the DY SFs for the \tauetauh channel in 2016 is $1.02\pm0.07$.

We measure the SF for the \wjets background in a CR defined to be orthogonal to the SR by inverting the \tauh selection, \ie, by requiring that the \tauh candidates fail the tight discrimination against jets while passing the very loose selection.
We also modify the $\mT>150\GeV$ SR requirement so as to include \PW boson production; we select events with $\mT < 190\GeV$, where the extended upper limit serves to probe the \wjets background in the high mass tail of the \mrec distribution. 
The contribution of QCD multijet events in this CR is nonnegligible, amounting to 30--40\% depending on the channel and data-taking year; it is estimated from data following the same methodology as that 
described below in this section for estimating the QCD multijet background in the SR.
In the \taumutauh channel, the SFs are $0.98\pm0.03$, $0.98\pm0.03$, and $0.92\pm0.03$ in 2016, 2017, and 2018, respectively. 
For the \tauetauh channel, the SFs are $1.02\pm0.07$, $0.94\pm0.08$, and $0.92\pm0.09$ in 2016, 2017, and 2018, respectively. 

For \ttbar events, we define the CR by imposing the SR selection criteria but changing the required number of \PQb jet candidates from zero to exactly one. 
The requirement of exactly one \PQb jet candidate ensures that the CR is kinematically similar to the SR. 
For 2016, 2017 and 2018, the SFs in the \taumutauh (\tauetauh) channel are $1.02\pm0.07$, $0.80\pm0.04$, and $0.82\pm0.04$ ($0.90\pm0.11$, $0.85\pm0.05$, and $0.85\pm0.05$), respectively. 
The difference in the scale factors across years is due to a known issue with track reconstruction during part of the 2016 data taking.

In all CRs we find agreement between data and simulation in the \mrec spectrum, 
after the corresponding SFs are applied to the simulation.

For the QCD multijet background estimate, we apply an ABCD method, in which four orthogonal regions are constructed with two measured observables. 
The estimate of the background yield $N_{\text{A}}$ in the SR is then determined from the yields in CRs B, C, and D as 
$N_{\text{A}} = (N_{\text{C}}/N_{\text{D}})N_{\text{B}}$. The necessary lack of correlation between the two observables has been validated in a separate 
ABCD set with $\ptmiss<30\GeV$ in the hadronic di-tau channel.
Here the first observable is the charge-sign pairing of the \PGt candidates: the OS requirement of the SR is reversed to same-sign (SS) to define CR B.  
The second observable is the appropriate measure of \PGt purity, namely lepton ID for \taul,  isolation for \tauh: both tight, as in the SR, 
for region B, while for regions C (OS) and D (SS) both must fail tight and pass loose ID. 
Contributions from background sources different from QCD have been removed in all CRs. 
The calculation is performed for each bin of the \mrec distribution, separately.

\subsection{The \texorpdfstring{\tauhtauh}{dihadronic} CRs}
\label{sec:bkgtauhtauh}

The main sources of background events in the \tauhtauh channel are DY and QCD multijet events. 

For the DY component, the SF is measured in a CR defined by the SR criteria except that the \tauh candidate pair's invariant mass $m(\tauh, \tauh)$ is required to satisfy $m(\tauh, \tauh) < 100\GeV$.
The DY SFs measured for the \tauhtauh channel are $1.11\pm0.10$, $0.96\pm0.09$, and $0.90\pm0.08$ in 2016, 2017, and 2018, respectively.

The QCD multijet background is estimated as follows in an ABCD approach:
A CR B is selected by the SR criteria except requiring an SS \tauh candidate pair. 
The common requirements for the events in the C and D CRs are $\ptmiss < 30\GeV$, and that both \tauh candidates fail the tight, while passing the very loose, discrimination against jets. 
Events entering CR C (D) pass the OS (SS) charge requirements.
To ensure that the sideband selection of passing the very loose and failing the tight $\tauh$ discriminator against jets in the low-\ptmiss sideband does not introduce any bias, the same method was also tested 
without the additional requirement of $\ptmiss < 30\GeV$, which was introduced to improve the QCD sample purity.
The results of this test, using the $\tauh$ sideband selection, were found to be statistically compatible with those obtained from the nominal selection, but with improved statistical precision in the sideband region. 
Therefore, the additional discriminator inversion method was adopted for the final result of the QCD estimate.

\subsection{The CR mass distributions}
\label{sec:bkgCRmass}

\begin{figure*}[!pht]
    \centering
    \includegraphics[width=.4\textwidth]{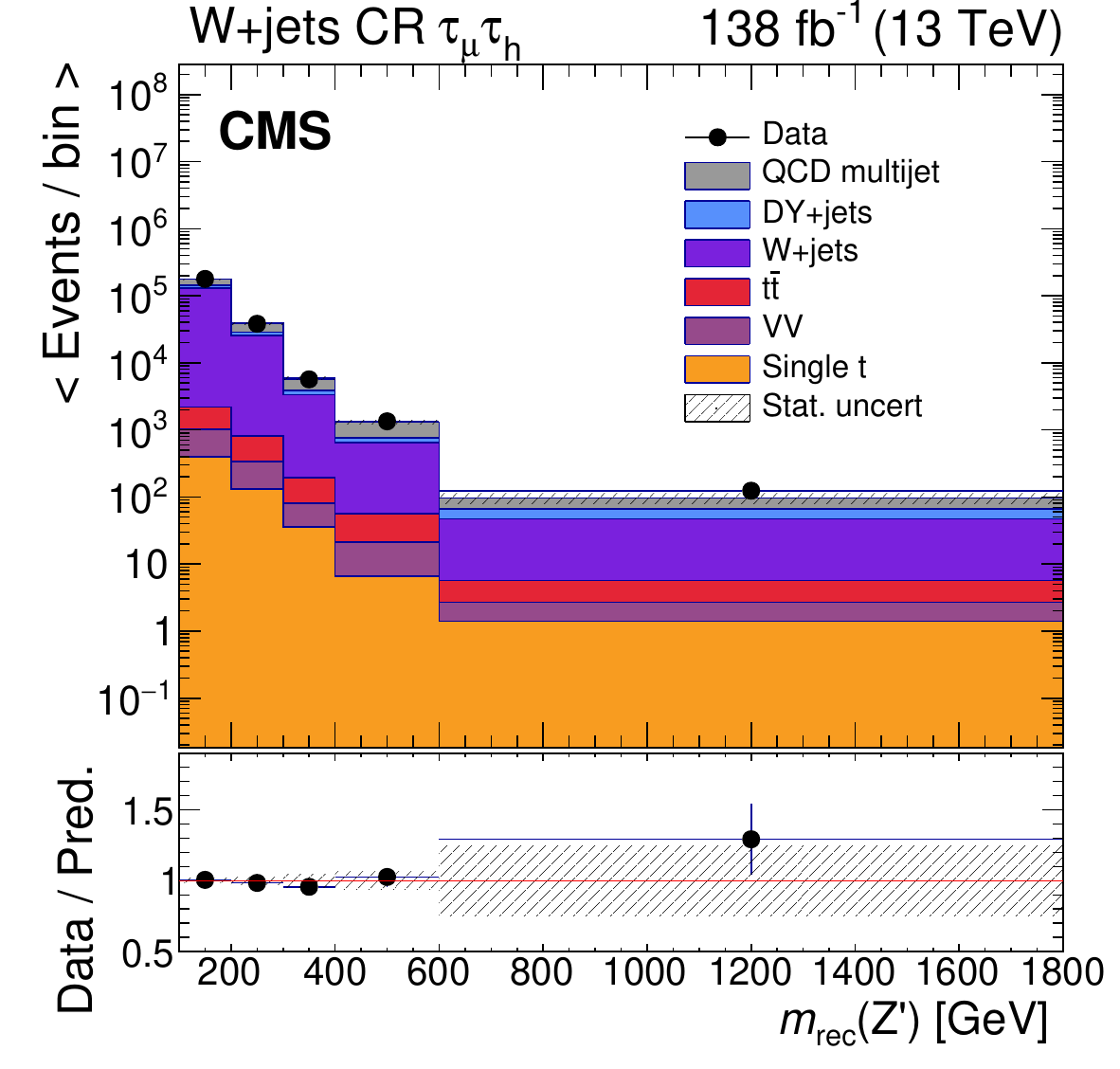}
    \includegraphics[width=.4\textwidth]{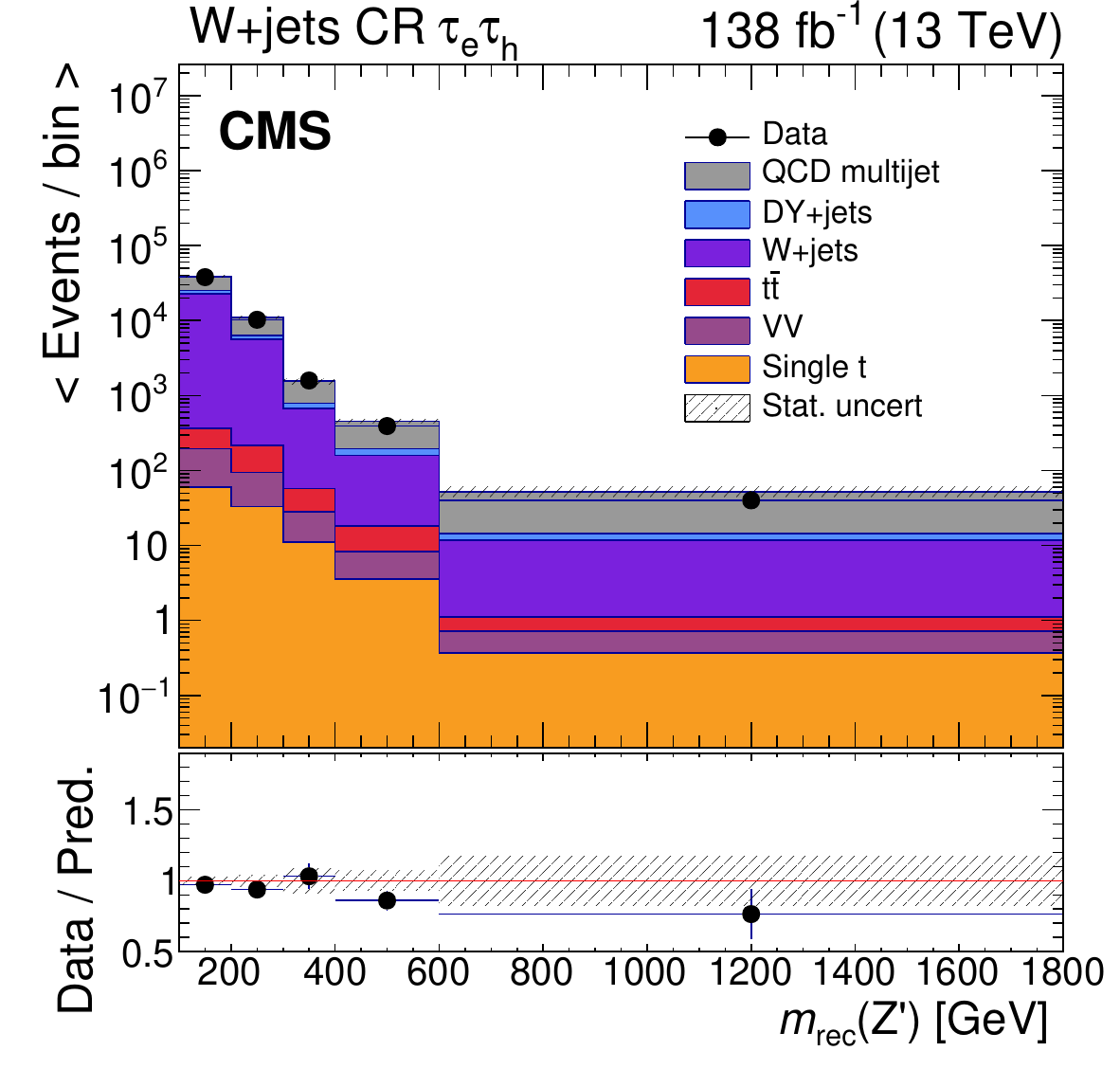}\\
    \includegraphics[width=.4\textwidth]{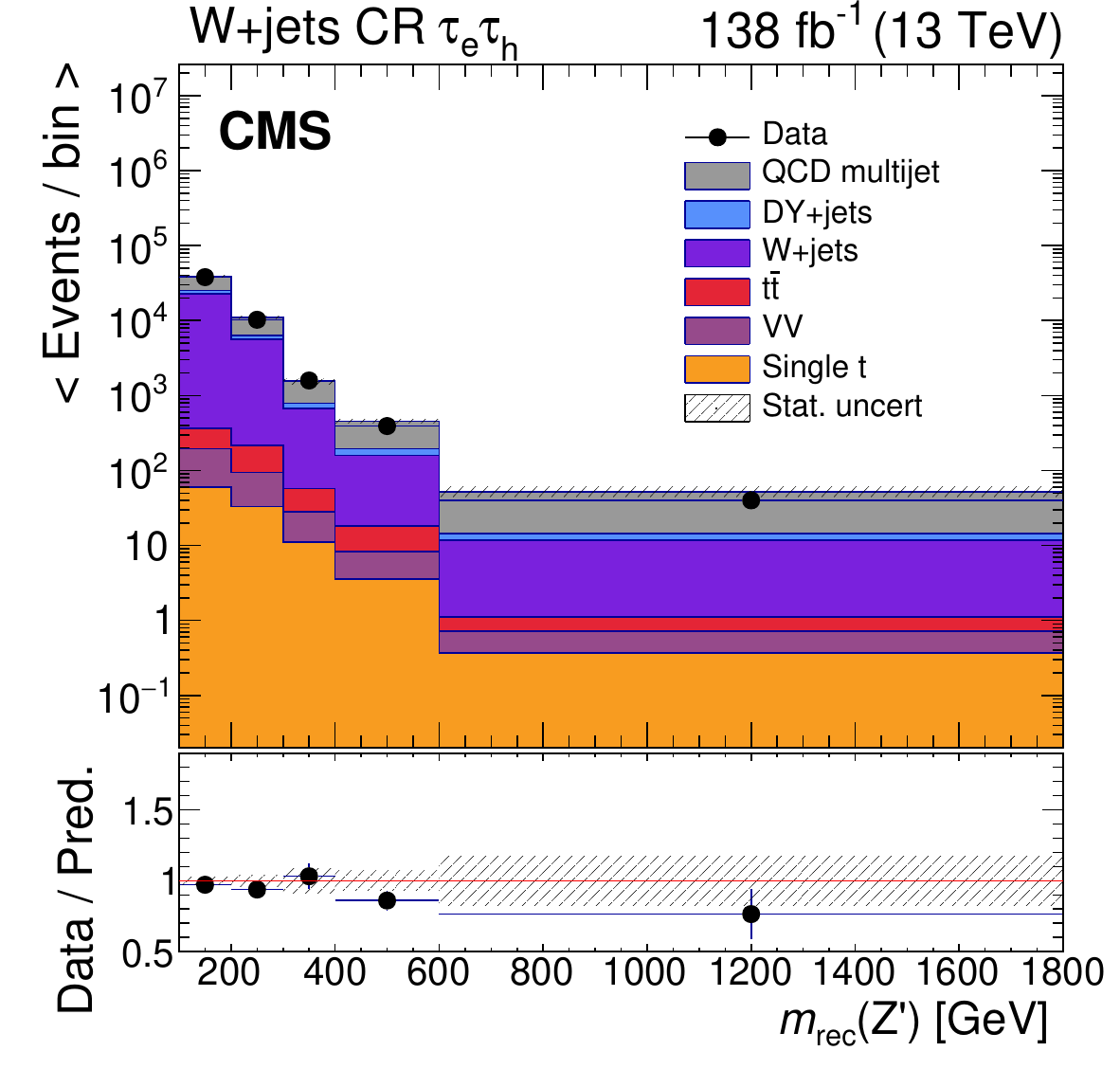}
    \includegraphics[width=.4\textwidth]{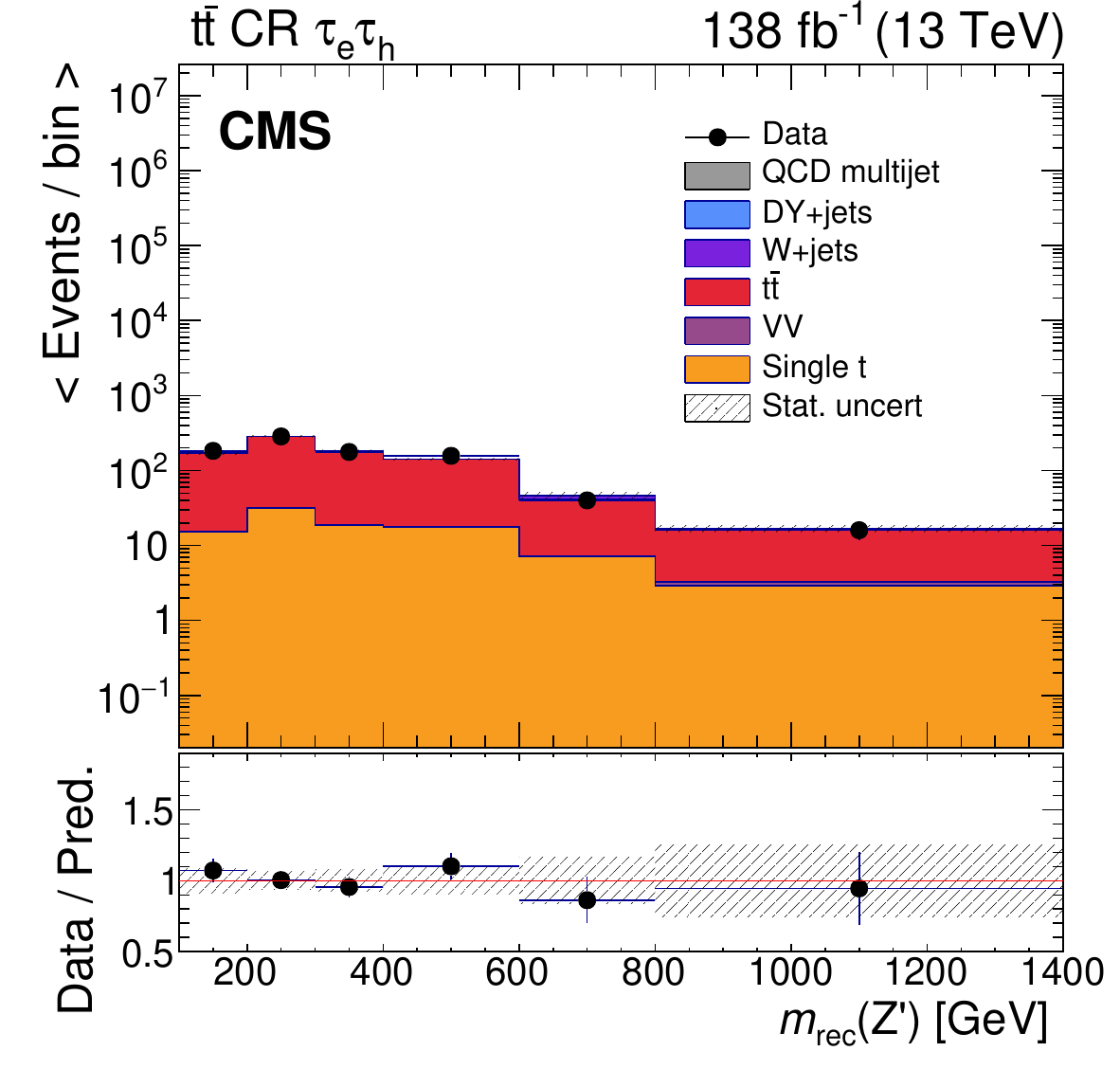}\\
    \includegraphics[width=.4\textwidth]{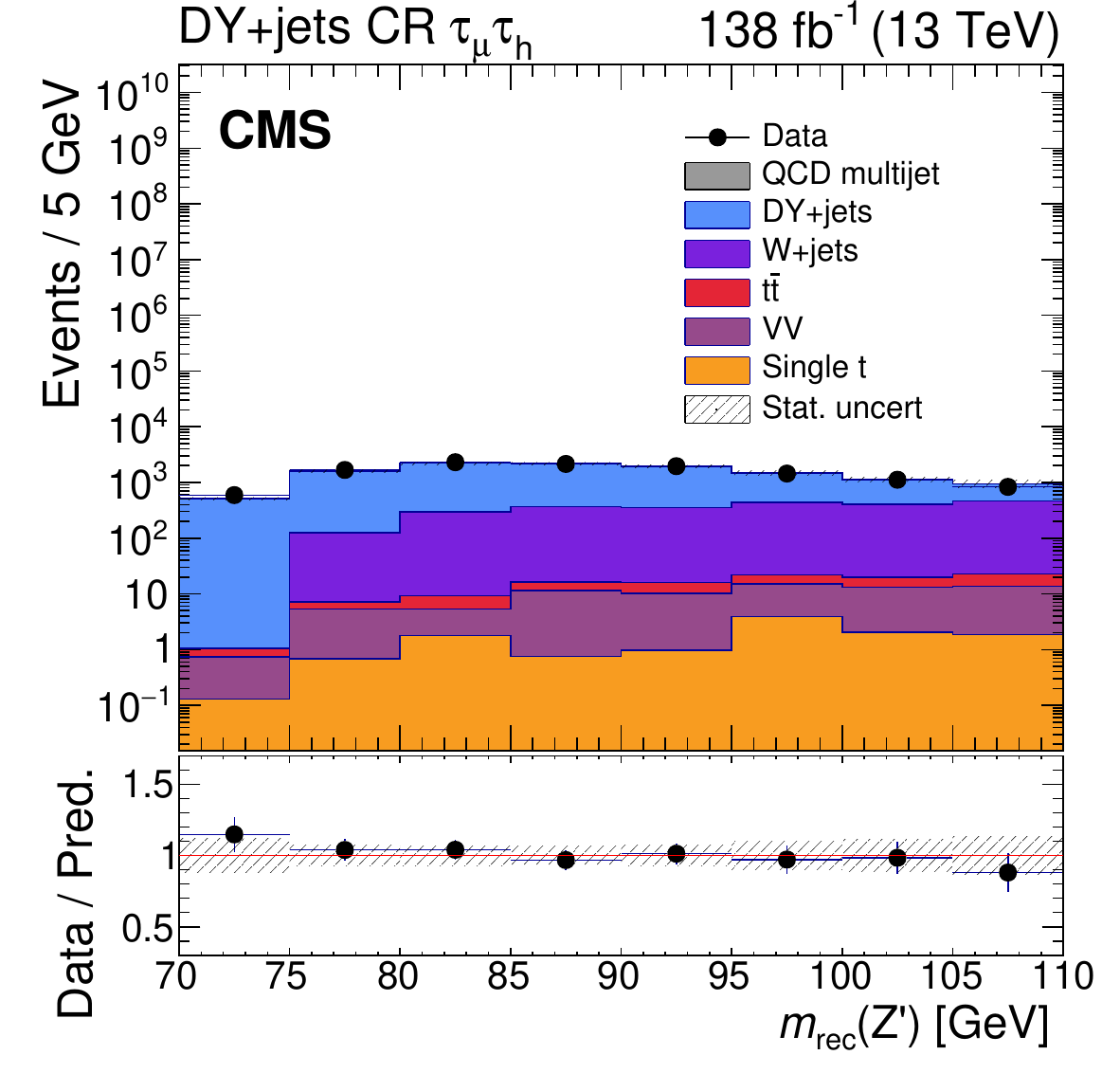}
    \includegraphics[width=.4\textwidth]{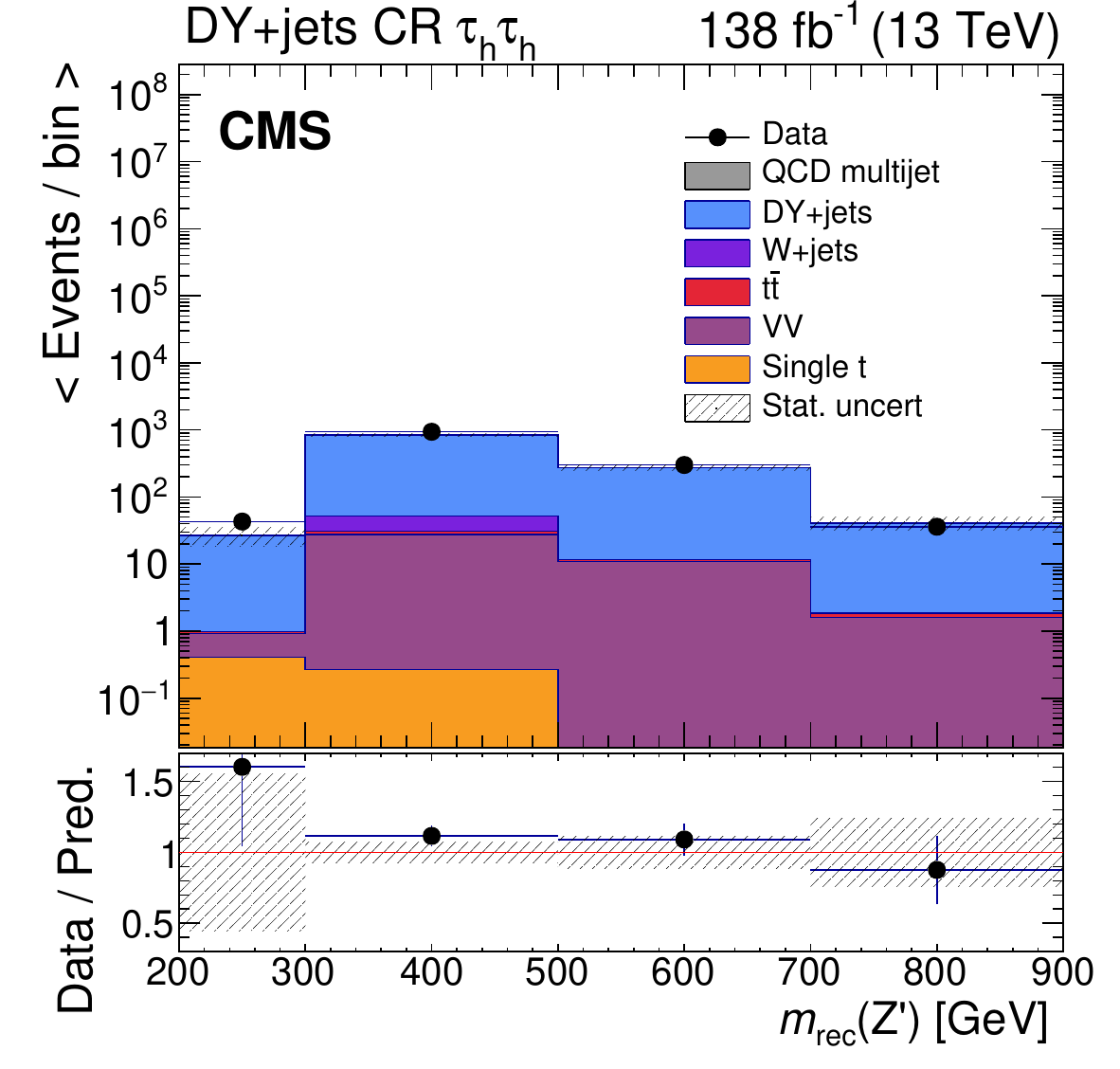}
	\caption{Pre-fit reconstructed mass distributions for the (upper row)
\wjets, (middle row) \ttbar, and (lower row) DY CRs, for the (left column) \taumutauh, (right column, upper and middle rows) \tauetauh, and (lower right)  \tauhtauh channel. 
	The targeted background in each CR, denoted by its CR name, has been corrected with the derived SF from that CR. 
	The background contributions appear as stacked solid histograms, while the data are shown as black solid markers with error bars. 
	Overflow counts are included in the last bin. 
	The gray shadow band represents the statistical uncertainty in the background prediction based on the number of simulated events and SF uncertainties. 
	The lower panel of each plot shows the ratio of data to predicted background yields.}
    \label{fig:bkgestmassplots}
\end{figure*}

Figure~\ref{fig:bkgestmassplots} shows the \mrec distributions for the data and simulation in the CRs that are used to compute the SFs.  Here the yields are summed over the data-taking years and are measured prior (pre-fit) to the ML fit that is used to extract the signal yields.

\section{Systematic uncertainties}
\label{sec:systematics}

We account for systematic uncertainties associated with the various sources as detailed in the following paragraphs.

The integrated luminosities for the 2016, 2017, and 2018 data-taking years have individual uncertainties of 1.2--2.5\%~\cite{CMS-LUM-17-003,CMS-PAS-LUM-17-004,CMS-PAS-LUM-18-002}, while the overall uncertainty for the 2016--2018 period is 1.6\%. 
This uncertainty is applied to all background and signal processes.

Uncertainties in the pileup modeling corrections are evaluated by varying the total inelastic \pp cross section used in the correction procedure by $\pm$5\%~\cite{CMS:2018mlc,ATLAS:2016ygv}.  The effect of the resulting change in the distribution of the number of primary vertices is less than 0.01\%, and is therefore treated as negligible. 

The efficiencies for the electron and muon reconstruction, ID, and isolation are measured with a ``tag-and-probe method"~\cite{CMS:2018rym,CMS:2020uim,CMS-DP-2020-021} with a resulting uncertainty of $\leq$2\%, depending on \pt and $\eta$. 
The total efficiency for the \tauh ID and isolation requirements is measured from a fit to the $\mu\tauh$ invariant mass distribution from $\PZ\to\tau\tau$ events, not taking 
into account neutrino hypotheses or \ptmiss. 
This is done in a sample selected with one isolated $\mu$ trigger candidate with $\pt>24\GeV$, and leads to a relative uncertainty of less than 9\% per genuine \tauh candidate~\cite{CMS:2018jrd}, dependent on \pt and $\eta$.
These uncertainties have contributions that are either statistical and thereby uncorrelated,  or systematic and therefore correlated across the data-taking years.
In the cases where background estimates rely on the modeling of the misidentification of jets as \tauh candidates, additional systematic uncertainties in the corresponding background yields are assigned 
that range from 20 to 30\%.

For the lepton momentum scale, the reconstructed momentum is shifted by 5\% and the \ptvecmiss vector recalculated.  
This results in a change in the expected signal yield of 1--6\%, depending on \mzp, and negligible change for the backgrounds. 
The corresponding resolution uncertainty in the simulation is taken as 0.03 times the difference between the reconstructed and true values of the lepton's momentum~\cite{CMS:2018rym}.  It is found to have negligible impact on either signal or background yields.

Similarly, we include the uncertainties in the energy scale and resolution of jets and the propagation thereof to \ptmiss~\cite{CMS:2016lmd}. The impact on the SR selection is indirect, affecting only the vetos on \PQb-tagged jets and VBF dijet candidates. 
Accordingly, the signal event yield is not affected by either uncertainty, while the background event yields vary, depending on the process, by less than 2\%.   

We propagate the uncertainty in the \PQb jet mistagging SF~\cite{BTV-16-002}
to its effect on the \PQb jet requirements in the SR selection.
The resulting uncertainty in the signal yield is found to be $\sim$1\%. 
The effect of the \PQb tagging efficiency uncertainty on the mass distribution in the \ttbar CRs for the \taultauh channels is measured by varying the \PQb tagging SF up and down by its uncertainty; the resulting uncertainty in the \ttbar SF is about 2\%.

The uncertainties in the cross sections used to predict the rates of various processes, which arise primarily from the factorization and renormalization scale choices and the PDFs, are propagated to the yield estimates. 
The cross section uncertainties are each separated into their scale (renormalization and factorization scales) and PDF components. 
They are correlated where appropriate among processes. 
To derive systematic uncertainties in the predicted signal yields arising from the choice of PDFs used in the simulation, we vary the PDFs following the PDF4LHC prescription~\cite{Butterworth:2015oua}. 
We find that these variations do not affect the shapes of kinematic distributions, but the resulting systematic uncertainty in the overall signal acceptance prediction is 2--6\%, depending on the assumed value of \mzp.

Uncertainties in the renormalization and factorization scales 
are evaluated as the differences in predicted signal yields
resulting from independent variation of these scales from their default values by factors
of 2.0 and 0.5~\cite{Cacciari:2003fi}.
The resulting uncertainties in the yields are $<$2\%, independent of \mzp.

The statistical uncertainty associated with the SFs measured in the CRs is propagated to the total uncertainty in the expected yields in the SR. 
The SFs in the \ttbar CRs, where some dependence on \mrec is observed, are fit with a first-order polynomial as a function of \mrec. 
The deviation of the resulting function from unity is taken as the systematic uncertainty in the shape.
The effect spans the range 1--22\%, depending on the \mrec bin and data-taking year.

For the QCD transfer factors $N_{\text{C}}/N_{\text{D}}$, the statistical uncertainties are in the ranges 19--23\% for \taumutauh, 23--26\% for \tauetauh, and 1.7--2.1\% for \tauhtauh. 
For each of the \taumutauh and \tauetauh channels, the QCD contribution to the SR is small (6--9\%), while it is large for the \tauhtauh channel (27--35\%), albeit mostly at low \mrec and therefore of low relevance to 
high-\mzp signal scenarios. Closure tests in a low-\ptmiss sideband version of the ABCD method for the QCD estimation in the \tauhtauh channel show agreement within statistical uncertainties, and so no additional closure systematic uncertainty is applied.
Instead, for all three channels, the transfer factors are fit with a first-order polynomial as a function of \mrec, and
the deviation of this function from unity provides the \mrec shape uncertainty.
The resulting systematic uncertainties for the three channels are 2.0--150\% (\taumutauh),
2.0--50\% (\tauetauh), and 0.2--24\% (\tauhtauh), depending on the data-taking year and mass value.
The large relative values of the shape systematic uncertainties occur at the high end of the \mrec distribution, especially for the \taumutauh channel, and arise from the limited statistical precision of the extrapolation.  Since the QCD multijet background is a very small fraction of the total background
in this \mrec region, the impact of this systematic uncertainty on the results is small.

\section{Results}
\label{sec:results}

The signal strength is extracted from an ML fit of the probability density function to the data yields in the SRs for all bins of the \mrec distribution, for all channels, and all data-taking years.  The probability density function is a sum of the corresponding predictions of the yields for the signal and for each of the SM background processes, with the signal strength as a free scale parameter modifying the signal prediction.
Statistical uncertainties from the CRs and simulated samples, as well as all systematic uncertainties, are introduced as nuisance parameters, including their correlations.  The nuisance parameters are constrained by log-normal penalty factors in the likelihood, which are uncorrelated across the \mrec bins. Any cross-contamination among the CRs is accounted for by the correlation terms between the CRs.  
The contamination of CRs by the signal is found to be small (order of percent or less), and is neglected in the fit model.   
The ML fit is implemented with the CMS statistical analysis tool \textsc{Combine}~\cite{CMS:2024onh}.

\begin{figure*}[!pht]
    \centering
	\includegraphics[width=.49\textwidth]{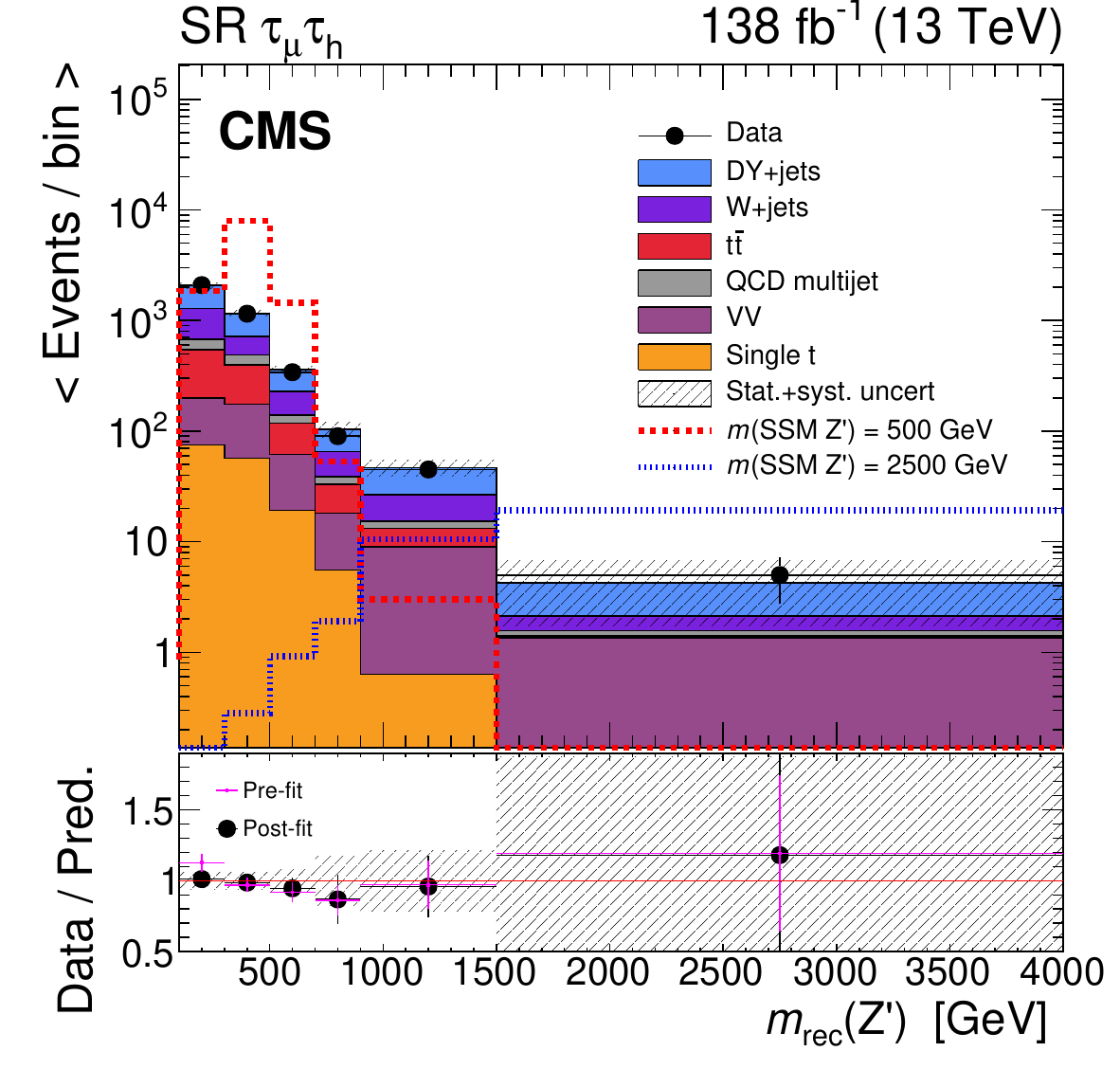}
	\includegraphics[width=.49\textwidth]{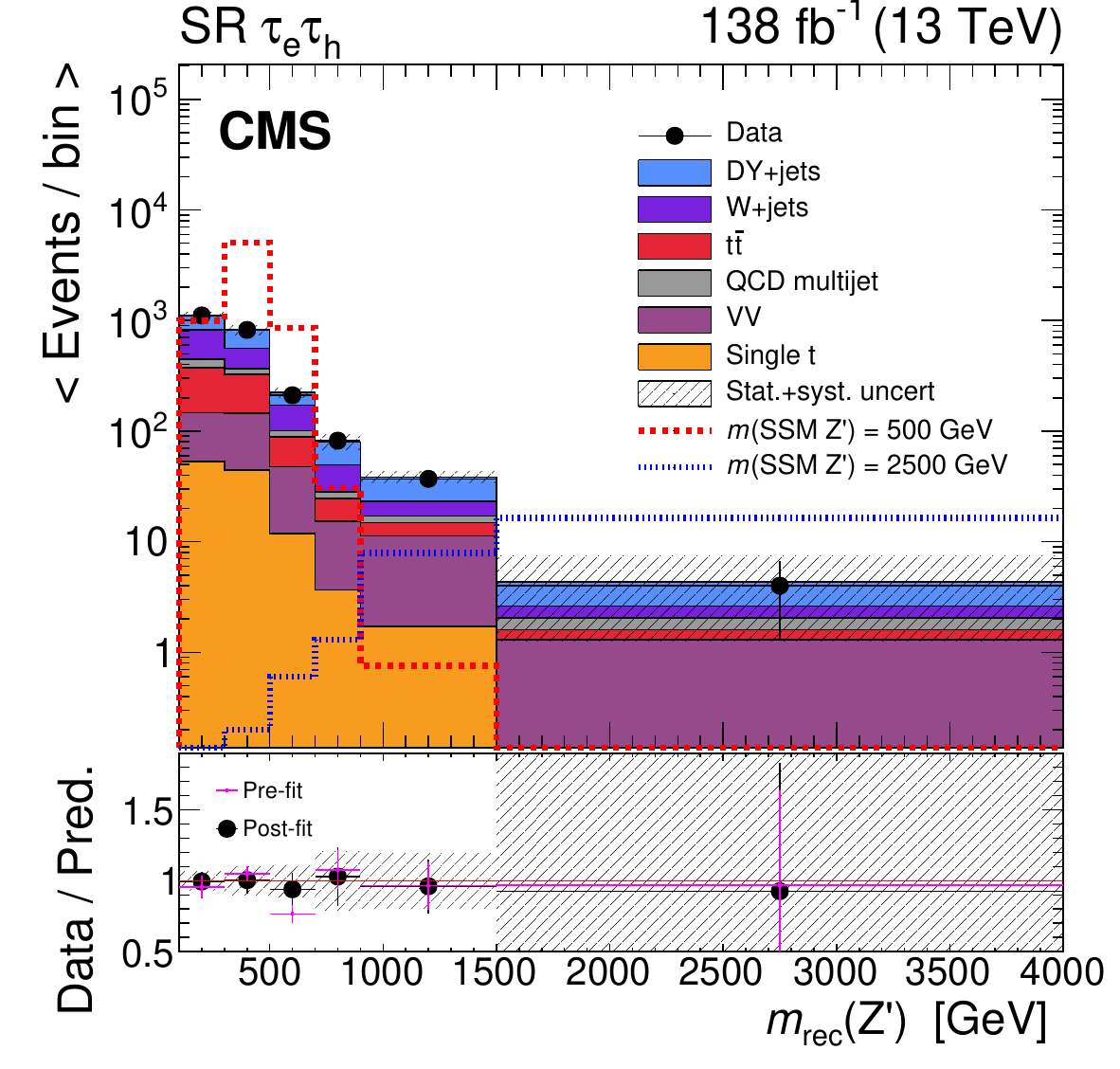}\\
	\includegraphics[width=.49\textwidth]{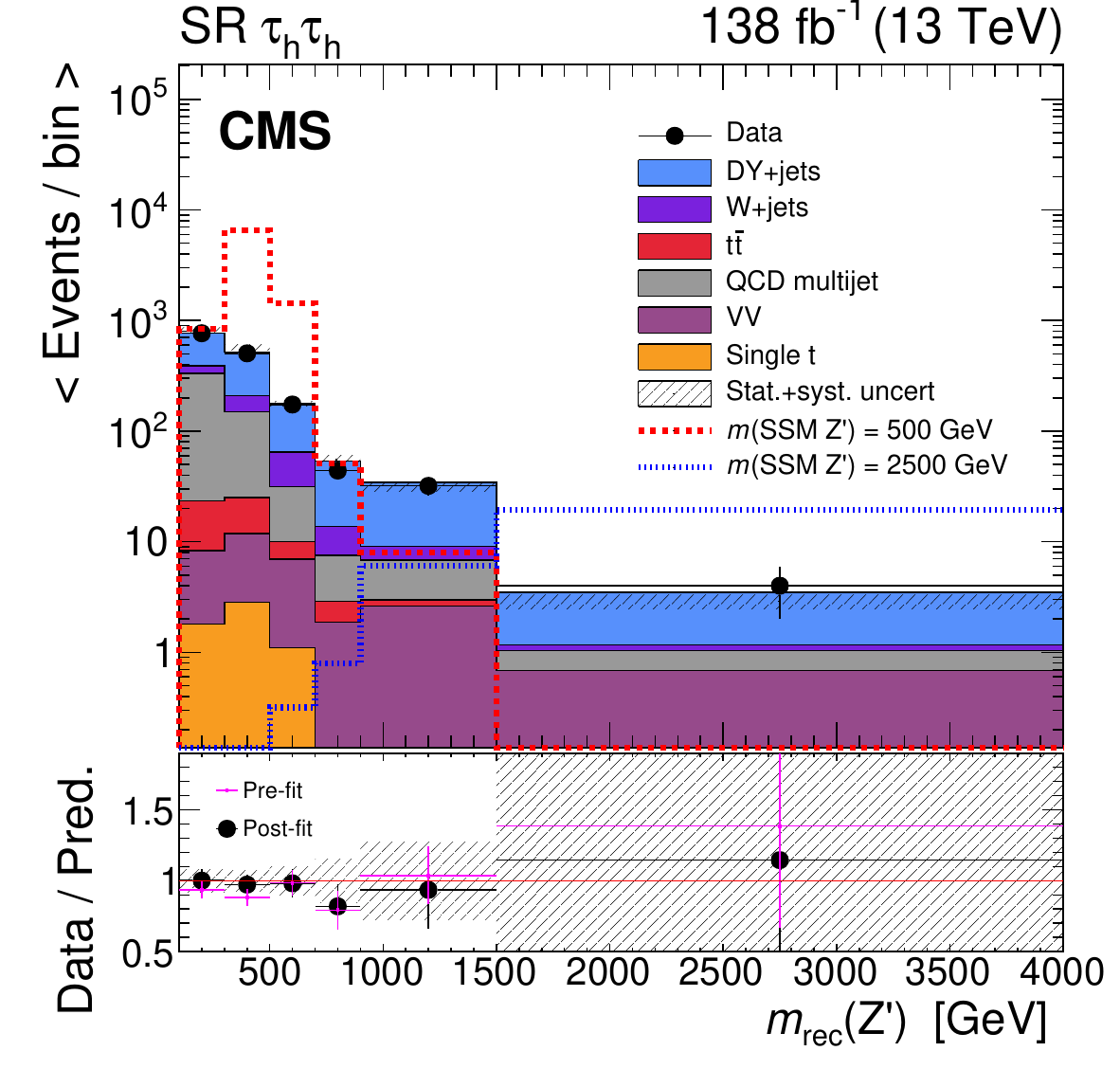}
    \caption{Post-fit reconstructed mass distributions for the (upper left) \taumutauh, (upper right) \tauetauh, and (lower) \tauhtauh channel.
  The background contributions appear as stacked solid histograms, while the observed data are shown as black solid markers with error bars.  Overflow counts are included in the last bin.  The gray shadow band represents the statistical plus systematic uncertainty in the post-fit background prediction. 
    The ratio to the pre-fit background prediction is indicated in magenta markers in the ratio.
        The signal expectation is shown in lines for \mzp of 500\GeV (red dashed) and 2500\GeV (blue dotted), normalized to 1850 and 0.131\unit{pb}, respectively.}
    \label{fig:massplots}
\end{figure*}

Table~\ref{tab:eventyields} shows for each channel the estimated pre-fit event yield for each SM background contribution, together with the total background and the observed yields.
The \mrec distributions for the data and the SM background estimates as determined by the background-only fits (post-fit) are displayed for the three channels in Fig.~\ref{fig:massplots}.  The lower panels compare the data with the pre-fit and post-fit estimates.
The observed yields are consistent with the SM prediction.

\begin{table*}[htb!]
\centering{
\topcaption{Pre-fit estimated background and observed event yields for the three channels in the SR.
        For each SM process the yield in events is shown,  along with the percentile contribution to the total.
        The background SFs derived from the dedicated CRs are used to determine the listed background yields.
        The uncertainty associated with the event yields is statistical. The uncertainty in the background contributions
        adjusted by the SFs includes the uncertainty in the corresponding SF. }
\begin{scotch}{lr@{\hskip 1pt}c@{\hskip 1pt}lrr@{\hskip 1pt}c@{\hskip 1pt}lrr@{\hskip 1pt}c@{\hskip 1pt}lr}
         & \multicolumn{4}{c}{\taumutauh} & \multicolumn{4}{c}{\tauetauh} & \multicolumn{4}{c}{\tauhtauh}  \\
Process                 & \multicolumn{3}{c}{Events}        & \%  & \multicolumn{3}{c}{Events}        & \%  & \multicolumn{3}{c}{Events}       & \%   \\
\hline
\wjets\rule{0pt}{2.2ex} & 851 & $\pm$ & 86    &  24 & 717& $\pm$ &89    &  30 & 158& $\pm$ &16   &  10    \\
DY                      & 1315& $\pm$ &70   &  37 & 661& $\pm$ &34    &  28 & 920& $\pm$ &43   &  55    \\
\ttbar                  & 637& $\pm$ &21    &  18 & 468.1& $\pm$ &8.1 &  20 & 32.8& $\pm$ &1.6 &   2    \\
Single \PQt             & 157.4& $\pm$ &6.0 &   5 & 116.1& $\pm$ &5.4 &   5 & 5.5& $\pm$ &1.2  & $<$1 \\
$\PV\PV$                & 302.4& $\pm$ &7.7 &   9 & 247.5& $\pm$ &6.0 &  11 & 26.8& $\pm$ &1.8 &   2    \\
QCD multijet            & 251& $\pm$ &44    &   7 & 135.0& $\pm$ &8.0 &   6 & 514& $\pm$ &27   &  31    \\[2pt]
Total background        & 3514& $\pm$ &122  & 100 & 2345& $\pm$ &96   & 100 & 1658& $\pm$ & 53 & 100    \\[2pt]
Data                    & \multicolumn{3}{c}{3737}          &     & \multicolumn{3}{c}{2269}          &     & \multicolumn{3}{c}{1549}         &        \\
\end{scotch}
\label{tab:eventyields}
}
\end{table*}

\begin{figure}[!ht]
	\centering
	\includegraphics[width=\cmsFigSpace]{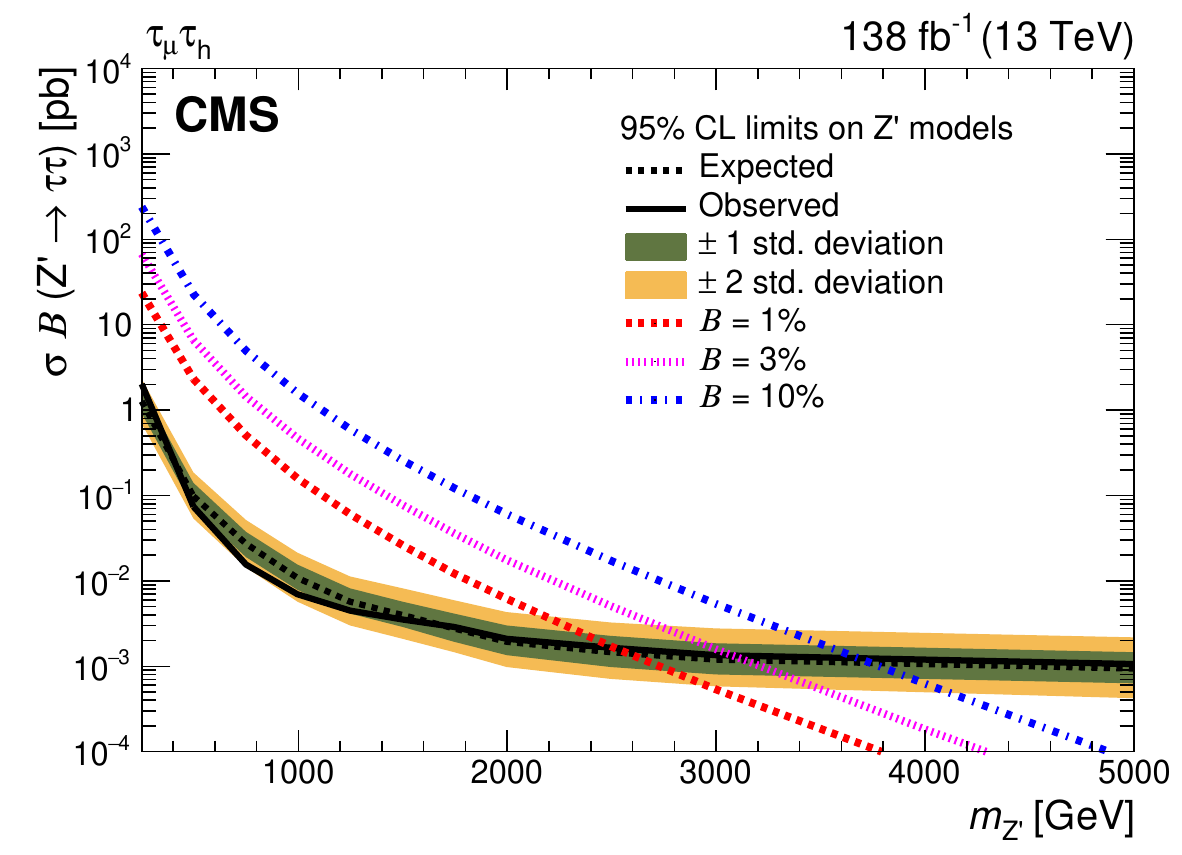}\\
	\includegraphics[width=\cmsFigSpace]{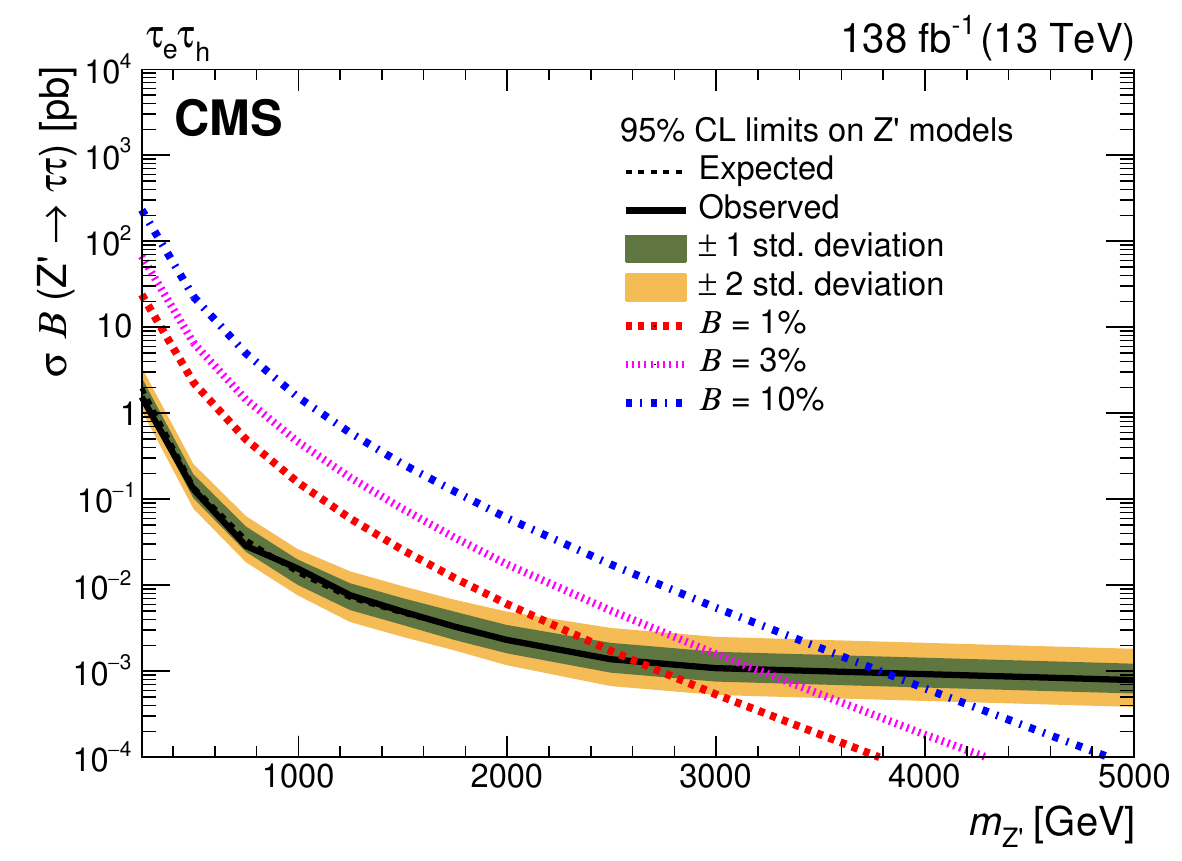}\\
	\includegraphics[width=\cmsFigSpace]{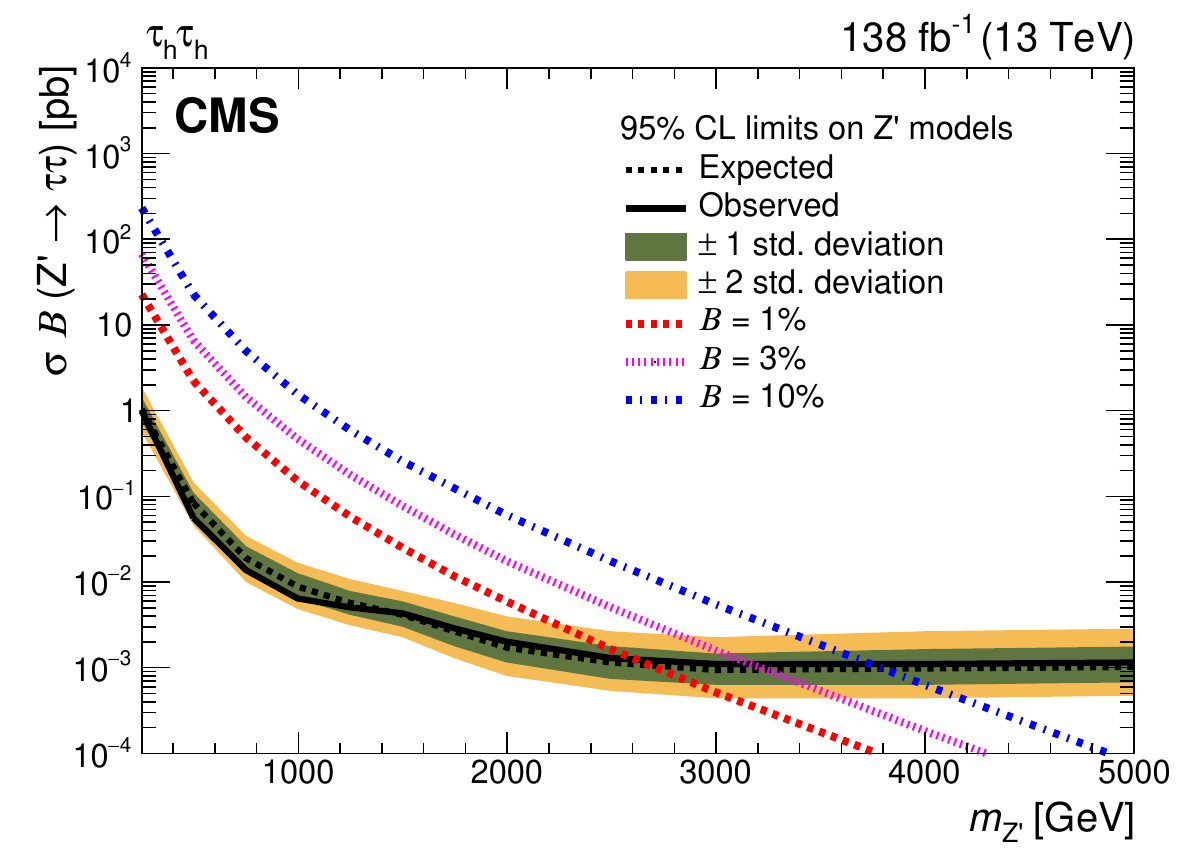}
	\caption{Upper limits at 95\% \CL on $\sigma\BZtautau$ for the (upper) \taumutauh, (middle) \tauetauh, and (lower) \tauhtauh channel. 
	In each plot, the solid black line represents the observed limit, and the black dashed line with green and yellow bands depicts the expected limit with its one- and two-standard deviation uncertainties, respectively. 
	The red dashed, magenta dotted, and blue dash-dotted lines represent the theoretical predictions for \BZtautau = 1, 3, and 10\%, respectively; 3\% is the value of \BZtautau in the SSM. These assume that the \BZtautau values are fixed for all \mzp.}
\label{fig:ChannelLimits}
\end{figure}

\begin{figure}[!ht]
	\centering
	\includegraphics[width=\cmsFigSpace]{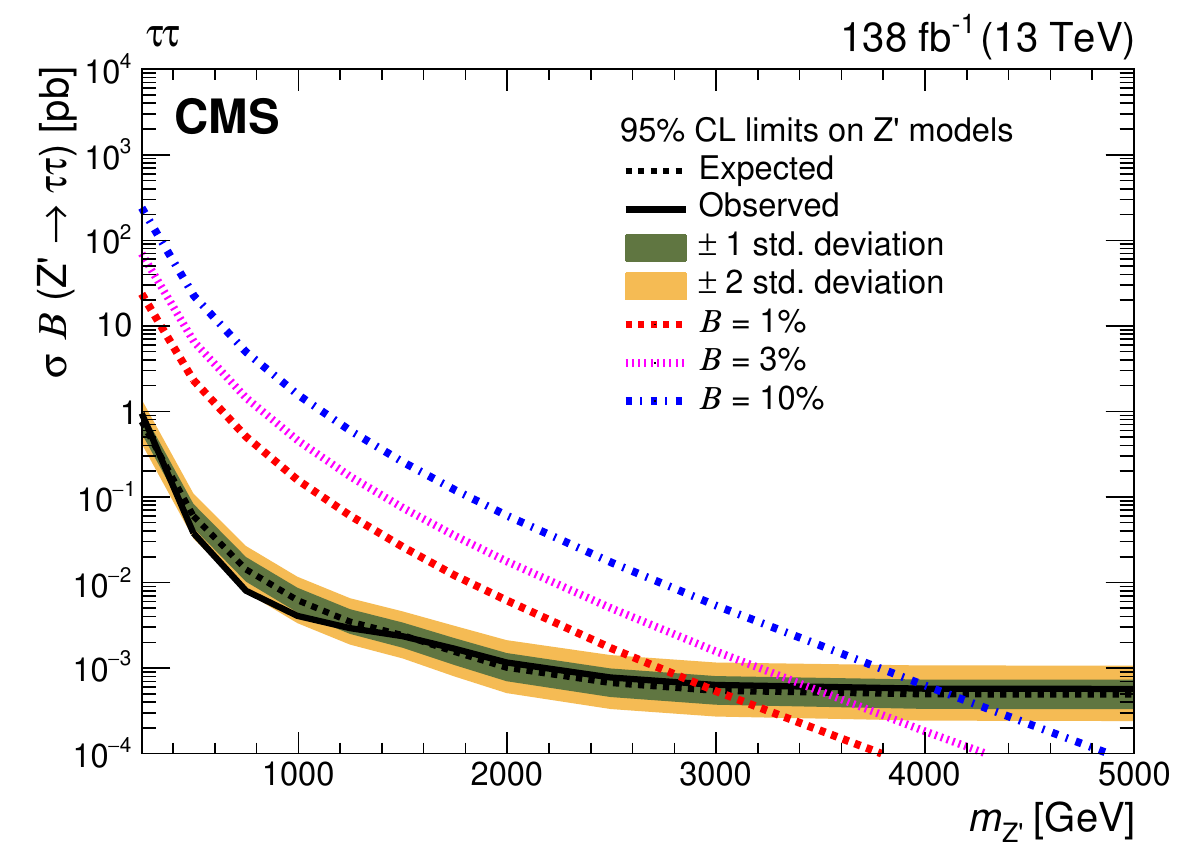}
	\caption{Upper limits at 95\% \CL on $\sigma\BZtautau$ for the combination of all channels.
	The solid black line represents the observed limit, and the black dashed line with green and yellow bands depicts the expected limit with its one- and two-standard deviation uncertainties, respectively.
        The red dashed, magenta dotted, and blue dash-dotted lines represent the theoretical predictions for \BZtautau = 1, 3, and 10\%, respectively; 3\% is the value of \BZtautau in the SSM. These assume that the \BZtautau values are fixed for all \mzp.}
	\label{fig:CombinedLimit}
\end{figure}

Using the \CLs criterion~\cite{Junk1999,bib-cls}, we compute an asymptotic approximation of the upper limit at 95\% \CL~\cite{Cowan:2010js}~on the product of the cross section and branching fraction $\sigma\BZtautau$, as a function of \mzp.
Figure~\ref{fig:ChannelLimits} shows the limits for each channel and Figure~\ref{fig:CombinedLimit} the combination of the three channels. 

Because of the large branching fractions of the tau lepton to hadronic
modes, the \tauhtauh channel provides the dominant contribution to
these limits across most of the \PZpr boson mass range. However, as
\mzp approaches 4-5 \TeV, there is a slight decrease in the \tauh
identification efficiency, which occurs because hadronic decays of Lorentz-boosted 
tau leptons lead to merged tracks and neutral hadrons. In
contrast, the identification efficiency for light leptons improves as
their energies increase to \TeV-scale values. As a result, for \mzp =
4-5 \TeV, the acceptance for the light lepton channels grows,
compensating for their lower branching fractions, and thus these
channels slightly outperform the \tauhtauh channel. There is no
significant difference in the impact between the mixed tau pair decay
channels.
The dashed lines represent the theoretical prediction for the branching fraction values of 1, 3, and 10\%, where 3\% is the approximate value of \BZtautau in the sequential standard model (SSM), considering \ttbar decays to be kinematically allowed.
The corresponding lower bounds on \mzp are 3.0, 3.5, and 4.1\TeV, respectively.

\section{Summary}
\label{sec:summary}
A search has been performed in proton-proton collisions at $\sqrt{s} = 13\TeV$ for the production via quark-antiquark annihilation of a heavy neutral gauge boson \PZpr decaying to $\PGtp\PGtm$.  The data were recorded at the LHC 
during 2016--2018, and correspond to an integrated luminosity of 138\fbinv.
As the tau lepton can decay hadronically (\tauh) or leptonically ($\PGt_{\Pell}$), with \Pell representing a muon 
or electron, the analysis includes the three decay channels with the highest \PGt lepton branching fraction products:
$\PZpr\to\taumu^{\pm}\tauh^{\mp}$, $\PZpr\to\taue^{\pm}\tauh^{\mp}$, and $\PZpr \to \tauh^{+} \tauh^{-}$.
An estimator \mrec is constructed to approximate the \PZpr boson mass \mzp in the presence of undetected neutrino daughters from the \PGt lepton decays.

The main background processes are estimated using Monte Carlo simulation adjusted by scale factors derived from data.
The observed yields are found to be consistent with the background prediction.  A shape-based analysis is performed using
the \mrec distribution as the fit discriminant to determine the likelihood for the observed data in the presence of a signal and the predicted background contributions.
The upper limit on the product of the \PZpr boson production cross section and decay branching fraction
is set at 95\% confidence level as a function of \mzp. 

In the scenario considered in this analysis, the data excludes a \PZpr boson with \mzp less than 3.0, 3.5, or 4.1\TeV for a branching fraction of 1, 3, or 10\%, respectively.
These exclusion limits are the most stringent to date for a \PZpr boson decaying to $\PGtp\PGtm$.

\begin{acknowledgments}
We congratulate our colleagues in the CERN accelerator departments for the excellent performance of the LHC and thank the technical and administrative staffs at CERN and at other CMS institutes for their contributions to the success of the CMS effort. In addition, we gratefully acknowledge the computing centres and personnel of the Worldwide LHC Computing Grid and other centres for delivering so effectively the computing infrastructure essential to our analyses. Finally, we acknowledge the enduring support for the construction and operation of the LHC, the CMS detector, and the supporting computing infrastructure provided by the following funding agencies: SC (Armenia), BMBWF and FWF (Austria); FNRS and FWO (Belgium); CNPq, CAPES, FAPERJ, FAPERGS, and FAPESP (Brazil); MES and BNSF (Bulgaria); CERN; CAS, MoST, and NSFC (China); MINCIENCIAS (Colombia); MSES and CSF (Croatia); RIF (Cyprus); SENESCYT (Ecuador); ERC PRG, RVTT3 and MoER TK202 (Estonia); Academy of Finland, MEC, and HIP (Finland); CEA and CNRS/IN2P3 (France); SRNSF (Georgia); BMBF, DFG, and HGF (Germany); GSRI (Greece); NKFIH (Hungary); DAE and DST (India); IPM (Iran); SFI (Ireland); INFN (Italy); MSIP and NRF (Republic of Korea); MES (Latvia); LMTLT (Lithuania); MOE and UM (Malaysia); BUAP, CINVESTAV, CONACYT, LNS, SEP, and UASLP-FAI (Mexico); MOS (Montenegro); MBIE (New Zealand); PAEC (Pakistan); MES and NSC (Poland); FCT (Portugal); MESTD (Serbia); MCIN/AEI and PCTI (Spain); MOSTR (Sri Lanka); Swiss Funding Agencies (Switzerland); MST (Taipei); MHESI and NSTDA (Thailand); TUBITAK and TENMAK (Turkey); NASU (Ukraine); STFC (United Kingdom); DOE and NSF (USA).

\hyphenation{Rachada-pisek} Individuals have received support from the Marie-Curie programme and the European Research Council and Horizon 2020 Grant, contract Nos.\ 675440, 724704, 752730, 758316, 765710, 824093, 101115353, 101002207, and COST Action CA16108 (European Union); the Leventis Foundation; the Alfred P.\ Sloan Foundation; the Alexander von Humboldt Foundation; the Science Committee, project no. 22rl-037 (Armenia); the Belgian Federal Science Policy Office; the Fonds pour la Formation \`a la Recherche dans l'Industrie et dans l'Agriculture (FRIA-Belgium); the F.R.S.-FNRS and FWO (Belgium) under the ``Excellence of Science -- EOS" -- be.h project n.\ 30820817; the Beijing Municipal Science \& Technology Commission, No. Z191100007219010 and Fundamental Research Funds for the Central Universities (China); the Ministry of Education, Youth and Sports (MEYS) of the Czech Republic; the Shota Rustaveli National Science Foundation, grant FR-22-985 (Georgia); the Deutsche Forschungsgemeinschaft (DFG), among others, under Germany's Excellence Strategy -- EXC 2121 ``Quantum Universe" -- 390833306, and under project number 400140256 - GRK2497; the Hellenic Foundation for Research and Innovation (HFRI), Project Number 2288 (Greece); the Hungarian Academy of Sciences, the New National Excellence Program - \'UNKP, the NKFIH research grants K 131991, K 133046, K 138136, K 143460, K 143477, K 146913, K 146914, K 147048, 2020-2.2.1-ED-2021-00181, TKP2021-NKTA-64, and 2021-4.1.2-NEMZ\_KI (Hungary); the Council of Science and Industrial Research, India; ICSC -- National Research Centre for High Performance Computing, Big Data and Quantum Computing and FAIR -- Future Artificial Intelligence Research, funded by the NextGenerationEU program (Italy); the Latvian Council of Science; the Ministry of Education and Science, project no. 2022/WK/14, and the National Science Center, contracts Opus 2021/41/B/ST2/01369 and 2021/43/B/ST2/01552 (Poland); the Funda\c{c}\~ao para a Ci\^encia e a Tecnologia, grant CEECIND/01334/2018 (Portugal); the National Priorities Research Program by Qatar National Research Fund; MCIN/AEI/10.13039/501100011033, ERDF ``a way of making Europe", and the Programa Estatal de Fomento de la Investigaci{\'o}n Cient{\'i}fica y T{\'e}cnica de Excelencia Mar\'{\i}a de Maeztu, grant MDM-2017-0765 and Programa Severo Ochoa del Principado de Asturias (Spain); the Chulalongkorn Academic into Its 2nd Century Project Advancement Project, and the National Science, Research and Innovation Fund via the Program Management Unit for Human Resources \& Institutional Development, Research and Innovation, grant B39G670016 (Thailand); the Kavli Foundation; the Nvidia Corporation; the SuperMicro Corporation; the Welch Foundation, contract C-1845; and the Weston Havens Foundation (USA).
\end{acknowledgments}
\bibliography{auto_generated} 

\cleardoublepage \appendix\section{The CMS Collaboration \label{app:collab}}\begin{sloppypar}\hyphenpenalty=5000\widowpenalty=500\clubpenalty=5000
\cmsinstitute{Yerevan Physics Institute, Yerevan, Armenia}
{\tolerance=6000
A.~Hayrapetyan, A.~Tumasyan\cmsAuthorMark{1}\cmsorcid{0009-0000-0684-6742}
\par}
\cmsinstitute{Institut f\"{u}r Hochenergiephysik, Vienna, Austria}
{\tolerance=6000
W.~Adam\cmsorcid{0000-0001-9099-4341}, J.W.~Andrejkovic, L.~Benato\cmsorcid{0000-0001-5135-7489}, T.~Bergauer\cmsorcid{0000-0002-5786-0293}, S.~Chatterjee\cmsorcid{0000-0003-2660-0349}, K.~Damanakis\cmsorcid{0000-0001-5389-2872}, M.~Dragicevic\cmsorcid{0000-0003-1967-6783}, P.S.~Hussain\cmsorcid{0000-0002-4825-5278}, M.~Jeitler\cmsAuthorMark{2}\cmsorcid{0000-0002-5141-9560}, N.~Krammer\cmsorcid{0000-0002-0548-0985}, A.~Li\cmsorcid{0000-0002-4547-116X}, D.~Liko\cmsorcid{0000-0002-3380-473X}, I.~Mikulec\cmsorcid{0000-0003-0385-2746}, J.~Schieck\cmsAuthorMark{2}\cmsorcid{0000-0002-1058-8093}, R.~Sch\"{o}fbeck\cmsAuthorMark{2}\cmsorcid{0000-0002-2332-8784}, D.~Schwarz\cmsorcid{0000-0002-3821-7331}, M.~Sonawane\cmsorcid{0000-0003-0510-7010}, W.~Waltenberger\cmsorcid{0000-0002-6215-7228}, C.-E.~Wulz\cmsAuthorMark{2}\cmsorcid{0000-0001-9226-5812}
% ...
\par}
\vskip\cmsinstskip
\dag:~Deceased\\
$^{1}$Also at Yerevan State University, Yerevan, Armenia\\
$^{2}$Also at TU Wien, Vienna, Austria\\
% ...
\end{sloppypar}
%%% END EDITABLE REGION %%%
% skeleton_end
\end{document}